          [2001/04/25 1.1 (PWD)]
\documentclass[a4paper,twocolumn]{esapub} 
\usepackage{natbib}
\usepackage{epsfig}
\usepackage{graphicx}

\title{\textit{INTEGRAL} and \textit{RXTE} observations of broad-line radio galaxy 3C 111}
\author[1,2]{M. Chernyakova}
\author[1,2]{P. Favre}
\author[1,2]{T.J.-L. Courvoisier}
\author[3]{A. Lutovinov}
\author[3]{S. Molkov}
\author[4]{V.~Beckmann}
\author[5]{A. Gros}
\author[4]{N. Gehrels}
\author[1]{N. Produit}
\author[1]{R. Walter}
\author[6]{A. Zdziarski}
\affil[1]{INTEGRAL Science Data Centre, Chemin d'Ecogia 16, 1290 Versoix, Switzerland}
\affil[2]{Geneva Observatory, ch. des Maillettes 51, CH-1290 Sauverny, Switzerland }
\affil[3]{Space Recearch Institute, Profsoyuznaya 84/32, 117810 Moscow, Russia}
\affil[4]{NASA Goddard Space Flight Center, Code 661,  Greenbelt, MD 
20771, USA}
\affil[5] {CEA Saclay, DSM/DAPNIA/SAp (CNRS FRE 2591), 91191, Gif--sur--Yvette Cedex,
France}
\affil[6]{ N. Copernicus Astronomical Center, Bartycka 18, 00-716 Warszawa, Poland}

\begin{document}

\def\deg{$^\circ$}
\keywords{X rays: radio galaxies; X rays: individuals: 3C 111}

\maketitle

\begin{abstract}
3C 111  is an X-ray bright broad-line radio galaxy
which is classified as a Fanaroff-Riley type II  source with 
a double-lobe/single jet morphology, and reported superluminal
motion. It is a well-known X-ray source, and was observed 
by every major X-ray observatory since \textit{HEAO-1}.  
In this paper we present the results of the \textit{RXTE} and \textit{INTEGRAL} data analysis
and compare them with the results of the previous observations.
\end{abstract}

\section{Introduction}

3C~111 ($z=0.0485$) is an X-ray bright broad-line radio galaxy (BLRG) 
which is classified as a Fanaroff-Riley type II  source with a double-lobe/single jet
morphology (Linfield \& Perley 1984, see Figure 1) and reported superluminal motion 
(Preuss et al. 1988).  It is  a well-known X-ray source, and was observed by every major 
X-ray observatory since \textit{HEAO-1}. 

\begin{figure}
\psfig{figure=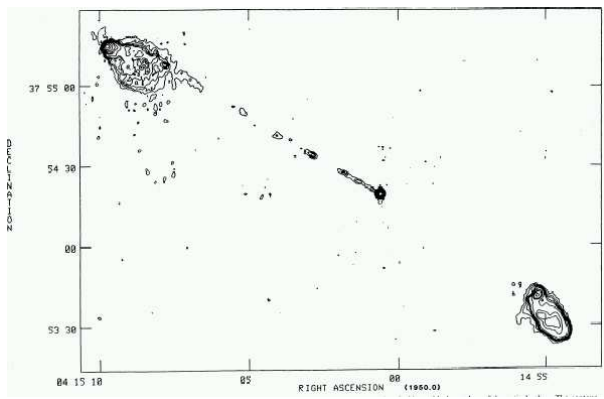,width=\columnwidth }
\caption{ Intensity map of 3C111 at 1.4 GHz. The bright central component is
 coincident with the nucleus of the  galaxy in the optical band.
 Figure taken from Linfield \& Perley 1984.}
\end{figure}

During the last decade numerous attempts were made to compare the X-ray properties 
of radio quiet  Seyfert~1 galaxies and  bright broad-line radio galaxies (BLRG) 
(e.g. Zdziarski et al. 1995, Wo\'zniak et al. 1998, Eracleous et al. 2000).
It was shown that the typical spectral shape of radio quiet  Seyfert~1 is
well described  in the 2--500 keV energy range by an intrinsic  power law with $\Gamma\sim 1.9$
with an exponential cutoff energy of the order of several hundred keV
and a Compton reflection component due to  reflection of the power-law photons from 
cold matter covering a solid angle of $\sim 2 \pi$  
(e.g. Nandra \& Pounds 1994,  Zdziarski et al. 1995).
As for the BLRG, their average spectrum, with a photon index
 $\Gamma\sim 1.7$ power law,  seems to have
no, or little indication of the reflected component. If associated with the jet their
 X-ray emission is likely to be related to non-thermal
Compton scattering.  With the present data a thermal origin of the  emission  still cannot
 be ruled out.

Main attention in all the X-ray observations of 3C~111 has been focused on characterizing
 its spectral shape. In the
 2 -- 20 keV energy range all measured spectra are consistent with a power-law spectrum modified
by the effects of neutral absorption (e.g. Weaver et al. 1995,
Nandra \& Pounds 1994, Reynolds et al.~1998).	 
 
For the hard X-ray spectrum we have less reliable results. Observations by \textit{HEXTE}, 
\textit{OSSE} and
\textit{BeppoSAX} are not sufficient enough to confirm or to rule out
 the break in the hard X-ray tail of 
the source spectrum (Wo\'zniak et al. 1998, Eracleous et al. 2000, Grandi et al. 2002). 

No significant short time variability was reported (see Eracleous et al. 2000 for a 
discussion on \textit{RXTE} 1997 data).
As for the long term
 variability, the 2-10 keV flux of 3C~111 varied by a factor of $\sim$5 during its whole observation
 history  (e.g. Reynolds et al.~1998). 

In this paper we present the results of the \textit{INTEGRAL}
 and \textit{RXTE} observations of 3C~111, 
and compare them with the
results of previous missions. We retrieved the {\it RXTE} data from the High Energy Astrophysics
Science Archive Research Center (HEASARC).The {\it INTEGRAL} and 
 part of the  \textit{RXTE} data presented here are 
 unpublished so far.

\section{INTEGRAL Observations}

The first time 3C~111 was in the \textit{INTEGRAL} field of view occurred during the
Galactic Plane Scan  observations  in March/April 2003. Unfortunately the source was too weak
to be significantly detected in 6.2 ksec during which the source had offset of less than 
10 degrees.

The second time \textit{INTEGRAL}  observed 3C~111 during  revolution
102 starting August 14 2003 for 57 ksec, during the calibration 
observations of the Crab. The observations were performed in 
staring mode, during which 3C~111 was  9\deg  off-axis.
Staring mode is the worst one for SPI, as in this case SPI data cannot be
used to reconstruct images and spectra of sources which are not on-axis
(Vedrenne et al. 2003).

Because of the offset of the source in both series of the
observations, 3C~111 was neither in \textit{INTEGRAL} X-ray monitor JEM-X, nor in 
the \textit{INTEGRAL} optical monitor OMC field of
view. The \textit{INTEGRAL} on-board imager IBIS (Ubertini et al. 2003)  
consists of two detectors, ISGRI and PICsIT, working in the energy band 
15 keV -- 10 MeV, and allows to localize sources with an accuracy down
to 30 arcseconds. The current version of the PICsIT data analysis is not working with 
more than eight degrees off-axis sources.
Thus in our analysis we were able to use only the data obtained by the
IBIS/ISGRI detector (Lebrun et al. 2003). Version 3.0 of the ISDC's
(Courvoisier et al. 2003) Offline
 Standard Analysis Software (OSA) was used.

\begin{figure*}
\centering
\psfig{figure=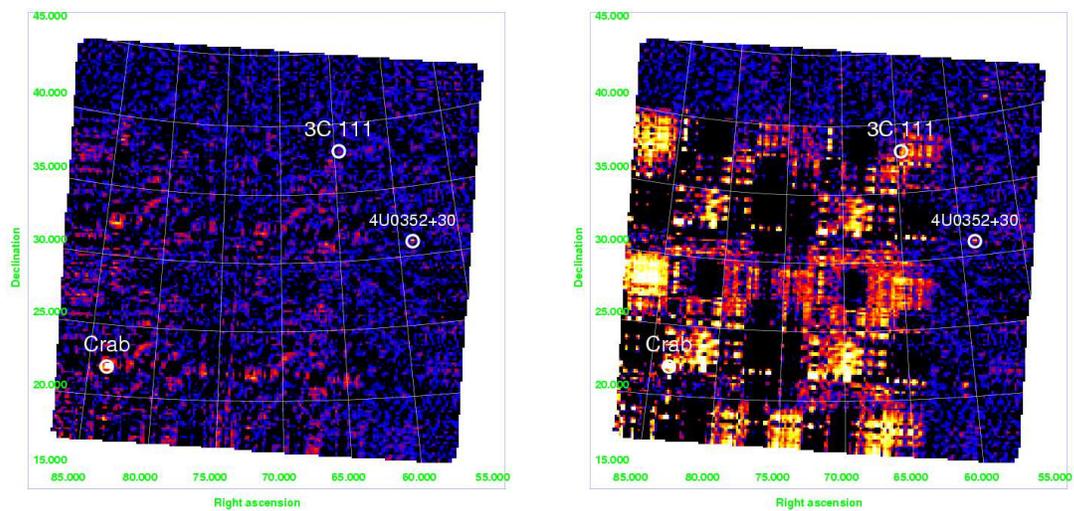,width=15cm }
\caption{ IBIS/ISGRI significance mosaic image in the 20--30 keV
energy band of the Crab region with 3C~111 in the field of view. Right panel shows the result
without ghost subtraction. 3C~111 is located in the region affected by the Crab.}
\end{figure*}
In  Figure 2  the August IBIS/ISGRI significance map of the Crab region
in  20-30 keV energy range is shown. It is a mosaic image of all the observations
with 3C~111 in the field of view, which results in  a total exposure of
57 ksec. The Crab is the brightest source of the region. Besides it,  4U 0352+30 (X-Persei)
is clearly detected at a significance level of 22.6 $\sigma$. 3C111 is detected 
at a significance level of 5.2 $\sigma$, the resulted form of the source is not round,
 and the source position
is only adjacent to  the brightest and most significant pixel (Figure 3). 

Such a behavior can be explained if we look at the right panel of Figure 2, in
which the image obtained without ghost subtraction is shown. Unfortunately 3C~111 
is located in the area affected by the Crab. Thus the only information that we can extract
from ISGRI data is the upper limit of the 3C111 flux, which is 9 mCrab in 20-30 keV,
3 mCrab in 30-40 keV and  6 mCrab in 40-60 keV energy ranges.

\begin{figure*}
\centering
\psfig{figure=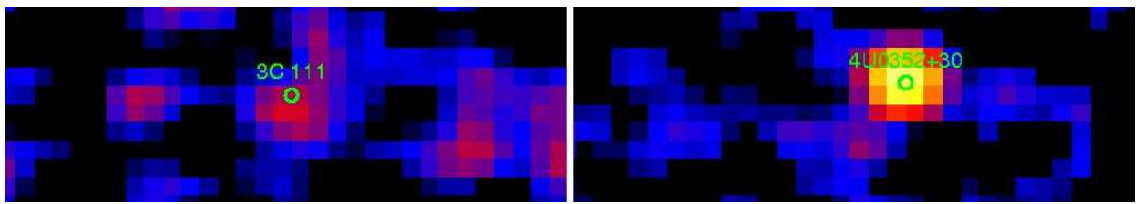, angle=-90,width=15cm}
\caption{ Enlarged region around 3C111 and 4U 0352+30.} 
\end{figure*}

\section{\textit{RXTE} Observations}
\textit{RXTE} public data are present in the HEASARC archive for 1997, 1999, 2001 and 2003.
 For the \textit{RXTE}/PCA data reduction
we used standard programs of the FTOOLS/LHEASOFT 5.2 package. 
For estimation of the background we applied the
"L7\_240"-based model. The results of the 1997 observations were
 already published  in the paper of
Eracleous et. al 2000.

\begin{figure*}
\centering
\psfig{figure=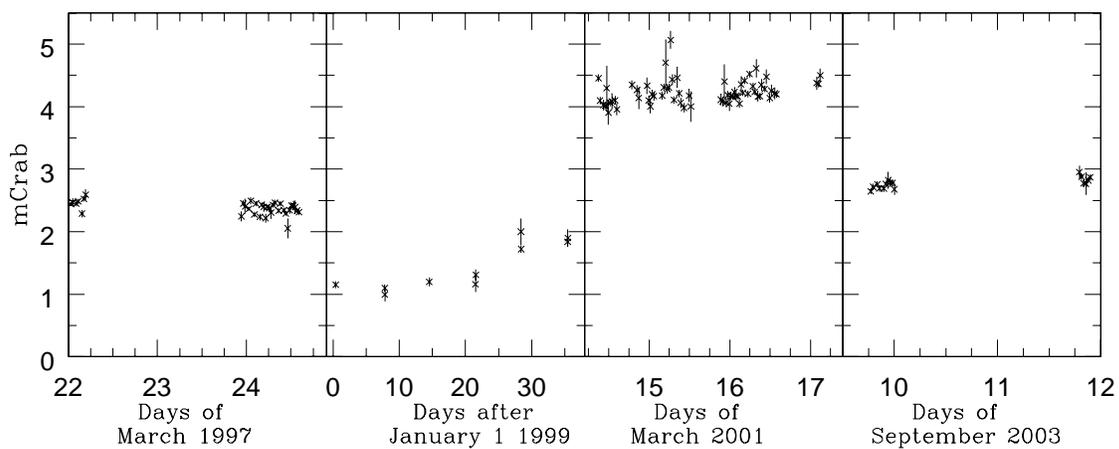, width=15cm}
\vspace{-2cm}
 \caption{ 3C~111 lightcurve, as observed by \textit{RXTE}/PCA. Time bins are of 1.8 ksec.} 
\end{figure*}

In  Figure 4 the lightcurve of the \textit{RXTE}/PCA observations of 3C 111 is presented.
No significant fluctuations on time scales less than an hour were observed. 
In 1999 the source was monitored
for more than a month. During this period the  source flux increased 
 from 1.0 to 1.8 mCrabs. There are also significant variations on the time scale 
of years. During
the \textit{RXTE}/PCA period of observations the intensity of the source varied from 1 mCrab in
 1999 to 4.5 mCrab in
2001. In 1997 and 2003 the source was at an intermediate level of 2.5 and 2.7 
mCrab, respectively.
The observed \textit{RXTE}/PCA flux in the 20-30 keV energy range is consistent with the upper 
limit observed by \textit{INTEGRAL}/ISGRI.

\begin{figure}[h]
\psfig{figure=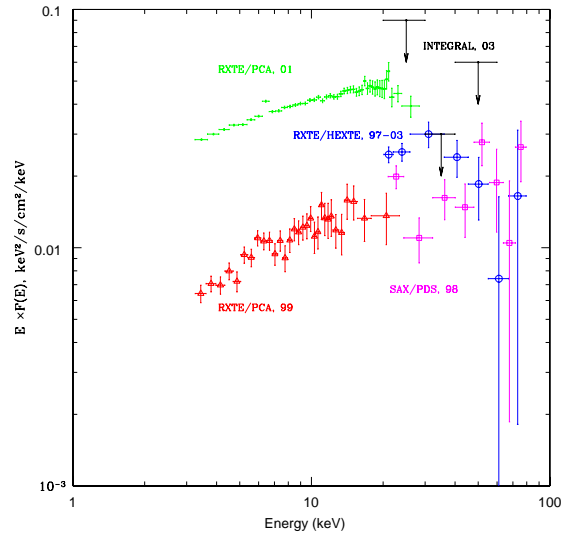, width=\columnwidth}
 \caption{ 3C~111 spectral history.} 
\end{figure}

In Figure 5 the PCA spectra of the brightest and faintest states of 3C111 during the \textit{RXTE} 
observations are shown  along with the average HEXTE spectrum and SAX/PDS spectrum.
In the  2 -- 20 keV energy range 3C~111 has typical for  BLRGs  spectrum, 
consistent with a power-law spectrum modified
by the effects of neutral absorption and Fe K$\alpha$   line emission. 
 Available high energy data do not
provide  information  whether the spectrum has a cutoff at higher energies.
 In Table 1 the best-fit
parameters for the spectra averaged over the given time period are listed. Only in
2001 emission line at 6.4 keV significantly improves the $\chi^2$ statistics. In 2003, due to 
problems with the response matrix, the spectrum is meaningful only above 6 keV.
\begin{table*}
\begin{center}
\caption{Spectral fitting results for \textit{RXTE} observations of 3C111}
\begin{tabular}{|l|c|l|l|l|l|l|}
\hline
Date & Energy  & Model& NH                & $\Gamma$& E$_{Fe}$&$\chi^2$/dof\\
     &Range, keV&     & $10^{22}$cm$^{-2}$&         & keV&\\  
\hline
03.1997&3 -- 25&NH*PL&3.3$\pm$0.15&1.9$\pm$0.01&--& 253/262\\ 
01.1999&3 -- 25&NH*PL&2.9$^{+0.8}_{-1.15}$&1.65$^{+0.05}_{-0.07}$&--& 502/309\\
03.2001&3 -- 25&NH*(PL+Gaus)&1$^{+0.1}_{-0.25}$&1.72$\pm$0.01&6.32$^{+0.12}_{-0.05}$&601/666\\
09.2003&6 -- 20&PL&--&1.62$\pm$0.4&--&110/64\\
\hline
\end{tabular} 
\end{center}
\end{table*}

 \section{Comparison with Previous Observations and Conclusions}

\begin{figure}[h]
\psfig{figure=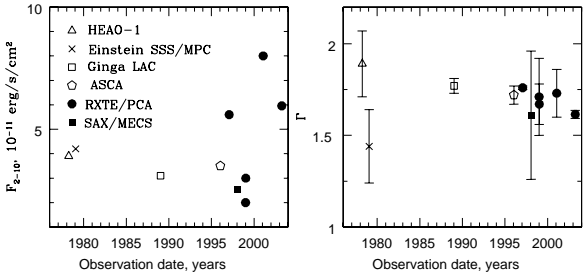, width=\columnwidth}
\vspace{-4cm}
\caption{ History of the 3C~111 flux in 2-10 keV energy range, 
and its power-law index variation.}
\end{figure}

In Figure 6 the  history of the 3C~111 flux in 2-10 keV energy range, 
and its power-law index variation is summarized. \textit{RXTE} was lucky to observe the source
during one of its faintest states (January 1999) and its brightest state (March, 2001).
However, no significant spectral changes are observed, except the fact that the brightest state
is the only one where the introduction of the Fe emission line significantly improves 
the $\chi^2$ statistics. But this effect can be easily explained by the higher data quality. 

\section*{Acknowledgments}
Authors are grateful to  M. Revnivtsev for the help in RXTE data reduction.
This work is based on the observations obtained 
through the \textit{INTEGRAL} Science Data
Centre (ISDC) and the High Energy Astrophysics 
Science Archive Research Center (HEASARC).
MC, SM and AL acknowledge International Space Science Institute
 for hospitality and financial support.

\end{document}